   \documentclass[prx,aps,preprint,superscriptaddress]{revtex4-1}

\usepackage{rotating}
\usepackage{times}
\usepackage{amsmath}
\usepackage{amssymb}
 
  \newcommand{\beq}{\begin{equation}}
\newcommand{\eeq}{\end{equation}}
\usepackage{graphics}
\usepackage{epsf}
\usepackage{amsmath}
\usepackage{amsfonts}
\usepackage{amssymb}
\usepackage{graphicx}
\usepackage{epsfig}
\usepackage{color}
\usepackage{bm}
\usepackage[colorlinks=true,citecolor=blue]{hyperref}
\usepackage{amsmath}
\usepackage{amssymb}
\usepackage{amsthm}
\usepackage{amsfonts}
\usepackage{times}

\usepackage{color}

\begin{document}


\title{  Physics of ultrathin films  and heterostructures  of rare earth nickelates }

\author{S. Middey}
\email{smiddey@uark.edu}
\affiliation{Department of Physics, University of Arkansas, Fayetteville, Arkansas, USA, 72701}
\author{J. Chakhalian}
\email{jchakhal@uark.edu}
\affiliation{Department of Physics, University of Arkansas, Fayetteville, Arkansas, USA, 72701}
\author{P. Mahadevan}
\email{priya@bose.res.in}
\affiliation{S. N. Bose National Centre for Basic Sciences,  Salt Lake, Kolkata, India, 700098}
\author{J. W. Freeland}
\email{freeland@anl.gov}
\affiliation{Advanced Photon Source, Argonne National Laboratory, Argonne, Illinois, USA, 60439}
\author{A. J. Millis}
\email{millis@phys.columbia.edu}
\affiliation{Department of Physics, Columbia University, New York, New York, USA, 10027}
\author{D. D. Sarma}
\email{sarma@sscu.iisc.ernet.in }
\affiliation{Solid State and Structural Chemistry Unit, Indian Institute of Science, Bangalore, India, 560012}


\begin{abstract}
The electronic structure of transition metal oxides featuring correlated electrons can be rationalized within the Zaanen-Sawatzky-Allen framework. Following a brief description of the 
present paradigms of electronic behavior, we focus on the  physics of rare earth nickelates as an archetype of complexity emerging within the charge transfer regime.Ê The intriguing 
prospect of realizing the physics of high $T_c$ cuprates through heterostructuring resulted in a massive endeavor to epitaxially stabilize these materials in ultra-thin form. A plethora of new phenomena unfolded in such artificial structures due to the effect of epitaxial strain, quantum confinement, and interfacial charge transfer. Here we review the present status of artificial rare-earth nickelates in an effort to uncover the interconnection between the electronic and magnetic behavior and the underlying crystal structure. We conclude by discussing future directions to disentangle the puzzle regarding the origin of the metal-insulator transition, the role of oxygen holes, and the true nature of the antiferromagnetic spin configuration in the ultra-thin limit.

\end{abstract}

\maketitle


\section{INTRODUCTION}

Complex oxides,  consisting of  transition metal  ions offer a vast landscape of fascinating phenomena  such as metal-insulator transitions, spin, charge and orbital orderings, high temperature superconductivity, colossal magnetoresistance and multiferroicity  etc~\cite{Imada:1998p1039}. The subtle competition between the bandwidth  ($W$), on-site electron-electron correlations ($U$),  crystal structure and charge transfer energy between oxygen and   transition metal sites ($\Delta$),  together with  the possibility of multiple   oxidation states typically available to transition metal ions and a wide range of structural variations  results in an unprecedented  diversity of interesting phenomena ~\cite{Zaanen:1985p418,Nimkar:1993p7355}.  The quest to understand and control these phenomena has captured the imagination of generations of condensed matter physicists and has fostered the development of  advanced  experimental probes  and theoretical methods. 

Thin films and heterostructures add an important new dimension to the problem.  Recent advances in  atomic precision oxide growth ~\cite{Zubko:2011p141,Hwang:2012p103,Chakhalian:2012p92, Chakhalian:2014p1189,Bhattacharya:2014p65} enable new classes of materials to be created and studied. Behaviors exhibited by ultra-thin films and heterointerface systems are different than in bulk, and the multiplicity of options for film growth and heterostructuring open new possibilities for control of material form and thus, for controlled comparison of the relation between structure and properties. For example, growing films  on different substrates lead to different strain states, and the materials respond with different degrees of octahedral tilts and rotations, resulting in different electronic properties. Similarly quantum confinement and interfacial charge transfer can change the underlying physics.   
  
In this review, we focus on the perovskite rare earth nickelate compounds, with chemical formula $RE$NiO$_3$ where $RE$ = La, Pr, Nd, Sm, Eu  ....Lu.  This material family exhibits rich physics including charge ordering, strong electron-lattice coupling, MIT and long period magnetically ordered state, all controlled by the interplay of crystal structure and electron correlations. Figure 1 schematically shows the rich diversity of ground state properties and their relationship to different physical parameters that can be controlled via diverse experimental realizations.

A crucial issue in the materials physics of complex oxides is  orbital control. In a simple ionic picture, the Ni is in  a low-spin $d^7$ state, with one electron in the two-fold degenerate $e_g$ orbitals. In bulk materials, the occupation probabilities of the two orbitals are nearly equal. An issue that has attracted some recent interest is the extent to which  ``orbital engineering''  (a controllable differential occupancy of the $e_g$ orbitals by changing strain and quantum confinement in ultra-thin films) is possible. The issue is important both in the context of the nickelate materials  and more generally as an example of control of many-body electronic structure. In this article, we review the known properties of the bulk $RE$NiO$_3$ materials and then how these are modified    in ultra-thin film  and heterostructured geometry.




\section{CLASSIFICATION OF CORRELATED MATERIALS FROM AN ELECTRONIC STRUCTURE POINT OF VIEW}
 
  \begin{figure*}[t!]
\includegraphics[width=6.0in]{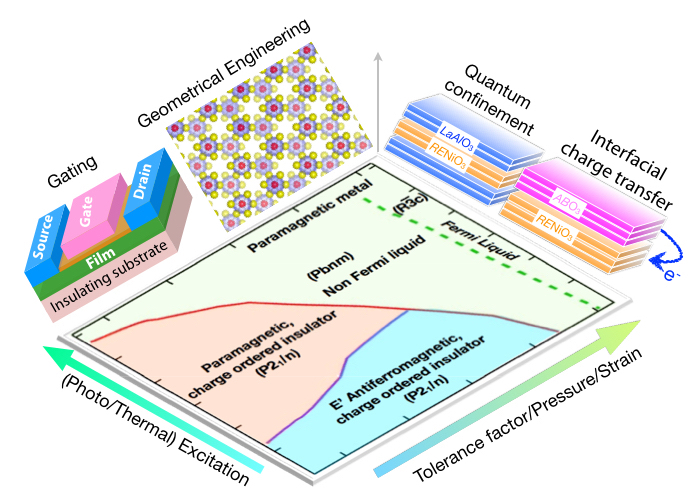}
\caption{Bulk phase diagram of  $RE$NiO$_3$ and various experimental ways
of altering these ground states, discussed in this review. }
\label{phasediagram}
\end{figure*}

As these $RE$NiO$_3$  are  correlated oxides, we begin this review with a basic description of electronic structure parameters of such transition metal compounds. Efforts to classify the electronic structure of transition metal compounds started with attempts to understand why many of these  have  insulating ground states  in spite of  having partially filled $d$ levels. Much attention has focused on NiO;  the notional electronic configuration of the Ni$^{2+}$ ion is 3$d^8$  in the crystal field split $t_{2g}^6 e_g^2$ configuration, with fully filled $t_{2g}$-derived bands and half filled $e_g$-derived bands. One of the early suggestions due to Slater~\cite{Slater:1951qv} emphasized the doubling of the unit cell due to the ground state antiferromagnetic order of NiO and the consequent opening of a gap in the half filled $e_g$ bands. A density functional theory (DFT)-based {\it ab initio} band structure calculation \cite{Terakura:1984p4734} indeed obtained a ground state insulator for NiO. However, this approach is not entirely satisfactory, because the calculated band gap value is an order of magnitude smaller than the experimental value and NiO continues to show  insulating behavior above the N\'{e}el temperature.

Mott~\cite{Mott:1949p416,Mott:1961p287} proposed an alternative idea that strong correlation effects in the $d$-shell can lead to an insulating phase even in  a system with partially occupied  levels. Hubbard analyzed a theoretical model based on Mott's ideas and showed \cite{Hubbard:1963p238,Hubbard:1964p237} that a sufficiently large on-site Coulomb interaction strength, $U$, within a partially occupied level can indeed localize electrons. The one-electron  removal spectrum, known as the lower Hubbard band (LHB)  [corresponding to the  occupied density of states (DOS) in a non-interacting system], is  separated from the one-electron-addition spectrum, known as the upper Hubbard band (UHB) [corresponding to the unoccupied part of the DOS], shown in {\bf Figure 2(a)}, when $U~>~U_{c}$, where $U_{c}$ is in the order of the  bandwidth ($W$) of the system dictated by the interatomic hopping  interaction strength, $t$. Such a system  known as the Mott-Hubbard insulator, and its   band gap is controlled primarily by the magnitude of $U$.


 Fujimori \& Minami's~\cite{PhysRevB:1984p957} analysis of spectroscopic features of NiO established the additional  necessity to account  explicitly for  oxygen 2$p$ levels, since the primary hopping interaction  $t_{pd}$  in NiO  connect O $p$ and Ni $d$ states. This requires the inclusion of an additional energy-scale, $\Delta$, needed to transfer an electron from the fully-filled oxygen levels to a Ni $3d$ orbital. The presence of the O $p$ band, arising from $p$-$p$ hopping interactions between different oxygen sites, in a typical transition metal oxide is shown in  {\bf Figure 2(a)} for $\Delta>U$ limit. In the opposite limit of $\Delta<U$, the energy ordering  of O $p$ states, LHB, and UHB are as shown in  {\bf Figure 2(b)} for charge-transfer insulators~\cite{Zaanen:1985p418},  with the O $p$ band appearing within the gap between UHB and LHB. In this  limit,  the band gap, measured between the top of the O $p$ band and the bottom of the UHB, is controlled more by $\Delta$ than by $U, $ in contrast to the case for Mott-Hubbard insulators ({\bf Figure 2(a)}). This scenario forms  the basis of the name, charge transfer  insulator, as  the lowest energy excitation for charge transport is achieved here by transferring an O $p$ electron to the transition metal (TM) $d$ orbital.

   \begin{figure*}
\includegraphics[width=6.5in]{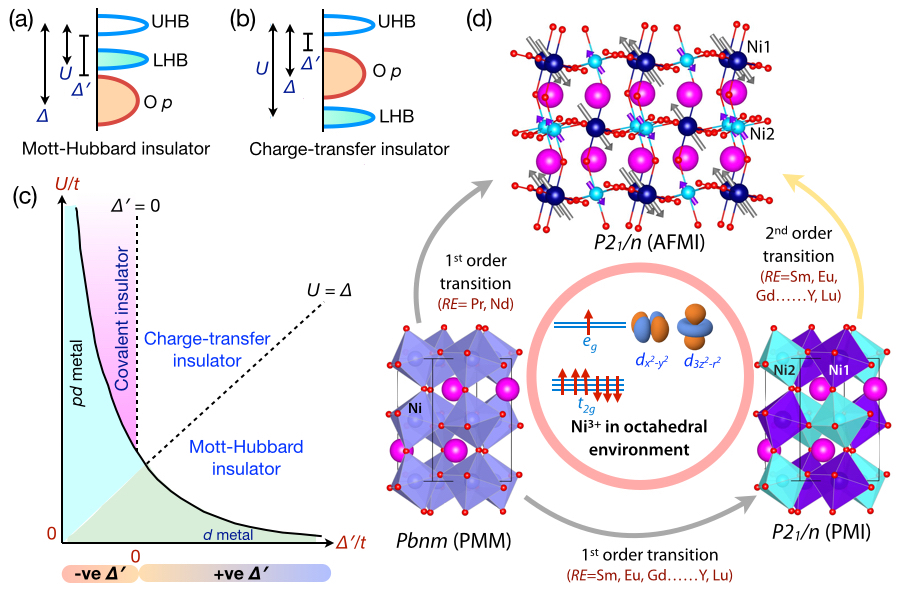}
\caption{ (a, b) Schematics of energy levels for (a) Mott-Hubbard insulator, (b) charge-transfer insulator. (c) Modified Zaanen-Sawatzky-Allen phase diagram~\cite{Zaanen:1985p418,Nimkar:1993p7355}.  (d) Various transitions observed in bulk $RE$NiO$_3$ series depending on the choice of $RE$  ion. The electronic configuration of a Ni$^{3+}$ ion is also shown for a pure ionic picture ($d^7$).}
\label{structure}
\end{figure*}

Combining these ideas, Zaanen, Sawatzky \& Allen \cite{Zaanen:1985p418} suggested that, instead of a single $U_{c}$ as envisaged by Mott and  Hubbard, the metallic and insulating ground states of transition metal compounds are separated by a line in the $U-\Delta$ phase space, with Mott-Hubbard insulators for $U_{c}<U<\Delta$ and charge transfer insulators for $U>\Delta>\Delta_{c}$. 3$d$ orbitals  contract across the $3d$ transition metal series, leading to a systematic  increase in $U$~\cite{Bandyopadhyay:1989p3517}. Within a related series of oxides, it is also  known ~\cite{Mahadevan:1996p11199} that $\Delta$ decreases across the Periodic Table due to  rapid stabilization of the $3d$ level with the increasing atomic number. This approach broadly categorizes insulating oxides of early 3$d$ transition metals as Mott-Hubbard type, with  larger $\Delta$ and  lower $U$, and those of late $3d$ transition metals as charge-transfer type, with smaller $\Delta$ and larger $U$. 
 
For sufficiently small $U$, the system becomes metallic, as in the Mott's description of the  insulator to metal transition, due to the overlap of the top of the LHB with the bottom of the  UHB. Similarly, one would expect that for a small enough $\Delta$ that makes the top of the  O $p$-band overlap the bottom of the UHB , defining the zero of the {\em effective} charge transfer energy, $\Delta '$ (see  {\bf Figure 2(b)}), the system will become metallic, as also suggested in the original Zaanen-Sawatzky-Allen phase diagram ~\cite{Zaanen:1985p418}. However, it was later argued \cite{Sarma:1990p45} that the insulating state will persist over a region  of $\Delta '~<~0$, depending on the value of $t_{pd}$ and $U$, becasue $t_{pd}$ will essentially mix the O $p$ and metal $d$ states, pushing them energetically apart to  form a nonzero bandgap, similar to the case of bonding-antibonding splitting formed in molecular  orbital theory. This new region in the $U-\Delta$ phase space was termed the covalent insulating regime \cite{Sarma:1990p45} to stress that  the band gap originated from  the presence of a sizable $t_{pd}$ mixing between O $p$ and metal $d$-states or covalency. This regime has also been referred to in the literature as the negative $\Delta$ insulator \cite{Mizokawa:1994p7193,Nimkar:1993p10927}, though care should be taken in retaining the distinction between $\Delta$ and the {\em effective} charge transfer energy, $\Delta '$, in this context. Explicit calculations \cite{Sarma:1992p531,Nimkar:1993p7355} showed that this covalent insulator regime is increasingly stabilized with increasing $t_{pd}$ and $U$.  {\bf Figure 2(c)} shows the resulting phase diagram, indicating different insulating and metallic regimes and establishing that the late transition metal oxides that are close to MITS, for example, many  interesting Ni and Cu compounds, including the nickelates discussed in this review,  belong to the covalent regime in their insulating  state. Increasing the formal valence of a transition metal ion has the primary effect of lowering $\Delta$ and increasing $t_{pd}$. Thus, compounds of earlier transition metals, such  as Co,  Fe, Mn and even Cr may also be in the covalent insulating regime in their higher valence states.
The correlation physics is different within the covalent insulating regime from the standard Mott paradigm due to the effective charge transfer energy being negative. One of the direct consequences of this negative effective charge transfer energy is that the ground state wave function has a significant and often dominant contribution from the $d^{n+1}\underline L$ state, where $n$ is the nominal \textit{d} electron count and  $\underline L$ denotes a hole state
generated in the ligand \textit{p} orbitals; this is in contrast to the Mott limit where the dominant contribution arise from $d^{n}$ states. In the case of nickelates, with a formal  $d^{7}$ configuration of  Ni$^{3+}$,  this ensures a significant contribution of  the $d^{8}\underline L$ state in the ground-state wave function as the expense of he contribution from  $d^{7}$ states.

The above discussion suggests that  localizing effects of increasing $U$ and $\Delta$ are obviously important; however, the delocalizing effects arising from various hopping interactions as well as any other  interaction  such as the crystal field and the spin-orbit interaction, that may influence  the bandwidth are also important and need to be treated on equal footing. This objective has been most successfully achieved through extensive developments of Dynamical Mean Field Theory (DMFT),  which provides the basis of most of our discussion of the electronic structure of nickelates  in later sections of this review. At a simpler level of broad classifications, we already know  from experimental studies that LaNiO$_3$ (LNO) is metallic, whereas all other members of the  series, $RE$NiO$_3$, where $RE$ is a rare-earth ion, are insulators at low temperature. Ab-initio electronic 
structure calculations for the series $RE$NiO$_3$ ($RE$=La, Pr, Nd, Sm and Ho)~\cite{Sarma:1994p10467}  established  a systematic reduction in the Ni $d$-bandwidth by approximately 10\% between La and Ho compounds. A fitting of the ab-initio band dispersions to a minimal tight-binding model with Ni $d$ and O $p$ states, did not show any substantial variation of $t_{pd}$ across the series. The bandwidth reduction can, therefore, be related to the reduction of the  hopping between different NiO$_6$ octahedra driven by a decreasing interoctahedral Ni-O-Ni angle with smaller $RE$ ions across the series. Various features in the valence band photoemission spectra of LaNiO$_3$ and NdNiO$_3$ (NNO) could be identified with features in the calculated density 
of states, underlining the importance of band structure effects\cite{Barman:1994p8475,Sarma:1995p1126}. An analysis of the $L_{23}M_{45}M_{45}$ Ni Auger spectrum revealed $U$ of Ni to be 4.7 $\pm 0.5$ eV. Hartree-Fock type mean-field solutions of a multiband Hubbard model  with these parameter strengths were found \cite{Barman:1994p8475} to be consistent with LNO being metallic and  NNO insulating.

\section{COMPLEXITY IN BULK \textit{RE}NiO$_3$}

Altough we begin our discussion of ultra-thin film nickelates by briefly reviewing several  key  phenomena present in the bulk  counterpart,   we refer the reader  to the reviews by Medarde~\cite{Medarde:1997p1679} and Catalan~\cite{Catalan:2008p729} these reviews for  more details concerning  bulk properties. Whereas the first member of the nickelate series, LaNiO$_3$ with rhombohedral $R\bar{3}c$ structure remains metallic down to the lowest temperature probed, the intermediate members with $RE$ = Pr and Nd undergo a first-order MIT from the paramagnetic metallic (PMM) state with orthorhombic ($Pbnm$) structure to an antiferromagnetic insulating (AFI) state accompanied with a lowering  of symmetry to monoclinic  ($P2_1/n$). As the  tolerance factor ($t$ = $d_{RE\mathrm{-O}}$/$\surd$2$d_\mathrm{Ni-O}$; where $d_{RE\mathrm{-O}}$ and $d_\mathrm{Ni-O}$ are ionic bond distances between \textit{RE} and O and between Ni and O respectively) decreases further away from unity for even  smaller  \textit{RE}   ions (e.g. Sm...Lu), the MIT\ temperature ($T_{\mathrm{MIT}}$) and magnetic transition  ($T_{N}$) start to separate  from each other ({\bf Figure ~1}).  Within the insulating phase,  the entire family has monoclinic  ($P2_1/n$) structure ({\bf Figure ~1 and 2(c)}) with two inequivalent Ni sites in the unit cell.  
 

 Electrical and thermal measurements on the bulk  and thin films revealed the signatures of complex multi-band behavior~~\cite{Ha:2013p125150}, which has been further corroborated by angle resolved photoemission spectroscopy (ARPES) measurement~\cite{Eguchi:2009p115122}. Surprisingly, despite   sharp changes in conductivity and lattice parameters across the MIT, optical conductivity measurements revealed only a  gradual  opening of band gap with   continuous  changes  in spectral weight of the Drude peak~\cite{Katsufuji:1995:p4830}. The reported optical  gap of $\sim$ 1 eV is also much larger than the  gap ($\sim$ 20 meV) obtained from  electrical conductivity data~\cite{Catalan:2000p7892}. Recent tunneling experiment showed the opening of a sharp gap ($\sim$ 30 meV) in the insulating state and also the existence of pseudo gap in metallic LaNiO$_3$~\cite{Allen:2015p062503}.
 
In the ionic picture, \textit{RE}NiO$_3$  compounds have  NiO$_6$ octahedra with Ni$^{3+}$ ions in the $d^7$ low-spin electronic configuration  ($t_{2g}^6 e_g^1$ : $S$ =1/2), which  are expected to be  Jahn-Teller active. However, resonant X-ray diffraction (RXD) experiments showed ordering of local magnetic moments on Ni sites in AFI phase and ruled out any orbital ordering~\cite{Scagnoli:2005p155111,Scagnoli:2006p100409}. The magnetic ground state of these spins  ({\bf Figure ~2(c)})  was initially deduced from neutron powder diffraction (NPD)~\cite{Garcia:1994p978}. The magnetic wave vector for the antiferromagnetic state was determined to be (1/2, 0, 1/2)$_{\mathrm{ortho}}$ [(1/4, 1/4, 1/4) in pseudo cubic notation]; the spin arrangement of this $E'$-AFM state can be viewed as either a sequence of $\uparrow \uparrow \downarrow \downarrow$   or $\uparrow \ \rightarrow \downarrow \leftarrow $ pseudo-cubic (1 1 1) planes and the non-collinear spin scenario  has been demonstrated by resonant x-ray scattering~\cite{Scagnoli:2008p115138}. Whereas two distinct  magnetic moments  1.4 $\mu_B$ and 0.7$\mu_B$ consistent with a checker board type charge ordering pattern were determined for  YNiO$_3$  ~\cite{Alonso:1999p3871},  the   magnetic moments in the  $RE$ = Pr~\cite{Garcia:1994p978,Vobornik:1999pR8426}, Nd~\cite{Garcia:1994p978,Vobornik:1999pR8426}, Sm~\cite{Carvajal:1998p456,Vobornik:1999pR8426}, Eu~\cite{Carvajal:1998p456,Vobornik:1999pR8426} compounds were  found to be similar on each Ni site. This surprising  result was  further corroborated by RXD experiments, in which  a very similar  energy dependence of magnetic scattering intensity  was recorded across the series, with an similar  spectroscopic signature for the local moment on Ni ~\cite{Bodenthin:2011p036002}. In view of the strong variation of structural distortion and the magnitude of $T_{\mathrm{MIT}}$  in the $RE$NiO$_3$ series, both measurements  raised  concern about  the true nature of the charge ordering state. A view which aligns many of these facts is the picture where the strong covalency leads to the ground state to be more $3d^8\underline{L}$, and the charge ordering pattern in this picture resides not on Ni but on the oxygen site (i.e.   $3d^8\underline{L}$ -  $3d^8\underline{L}$ $\rightarrow$  $3d^8$ -  $3d^8\underline{L}^2$)~\cite{Mizokawa00}.

\section{EPITAXIAL STABILIZATION }
 The +2 charge state is the most common oxidation state of Ni and the    \textit{RE}NiO$_3$ requires Ni oxidation state to be +3.  Such high oxidation state is stabilized only under a  high oxygen pressure \textit{P}  $\sim 200$ bar, and an elevated temperature of $1000^\circ $C~\cite{Lacorre:1991p225}.  Despite the extreme synthesis conditions, to-date only micrometer-sized single crystals have been obtained  for these nickelates (except for  LNO) ~\cite{Alonso:2004p1277}, limiting the application of many experimental techniques crucial for understanding the underlying physics of these materials. As an alternate route,   the growth of thin films is a feasible solution for obtaining large-size single-crystalline materials. Not  surprisingly,  many crucial pieces of information  discussed in the last  section were obtained by using single-crystalline thin films. 

\subsection{Thermodynamic Stability:} 
Whereas thin films of LNO could be grown without  high oxygen pressure by  pulsed laser deposition (PLD)~\cite{Prasad:1993p1898} and sputtering~\cite{Yang:1995p2643},    stoichimetric NdNiO$_3$  films with  a bulk-like MIT were initially achieved only  by postannealing under a high oxygen pressure ( $\sim$115 bar)~\cite{DeNatale:1995p2992}. Later,  epitaxial films for several members of the family  were successfully synthesized    at  much reduced oxygen pressure ($<$ 0.02 bar) by  metal-organic chemical vapor deposition (MOCVD) techniques by Novojilov et al.~\cite{Novojilov:2000p2041}. Following this success, the high  quality epitaxy at low oxygen partial pressure has been also realized  by  PLD~\cite{Catalan:2000p606}, sputtering~\cite{Son:2010p062114} and all-oxide molecular beam epitaxy (MBE)~\cite{Nikolaev:1999p118}. 

To understand how a single crystalline phase can be obtained outside  the thermodynamic phase diagram $[P(O_2$)-$T]$ of bulk materials, we consider the model of epitaxial stabilization of thin films~\cite{Kaul:2004p861}. In this phenomenological framework, the relative difference between the free energies for the  bulk and epitaxially stabilized phases is given by:
\begin{equation}
\Delta E = \Delta E^{ic} -  \Delta E^{c} = h [\Delta g^{ic}_v - \Delta g^{c}_v -\frac{\mu}{1-\nu}\epsilon^2] + [\sigma^{ic}_s - \sigma^{c}_s] \ 
\end{equation}
Here the  superscripts   ${c}$ and ${ic}$ denote the epitaxially coherent and incoherent (i.e. free from substrate) phases, respectively, $h$ is  the film thickness, $\Delta g_v$ is the Gibbs free energy per unit volume, $\mu$ is the shear modulus, $\nu$ is a  Possion's ratio, $\epsilon$ is the relative lattice mismatch between the unit cell parameters of the desired phase and the substrate and $\sigma_s$ is the surface tension. Whereas the volume contribution (the first term) is always negative, the surface contribution (the second term) is always positive. If the energy contributions are such that $\Delta E>$ 0, then an epitaxially coherent phase is stabilized\ \cite{Kaul:2004p861}.  As  seen in equation 1, a dramatic reduction in the contribution to the free energy comes from the coherent film-substrate interface, implying that  $RE$NiO$_3$ phases can be stabilized  on a perovskite substrate whereas the films on non-perovskite substrates like MgO and ZrO$_2$ lack the required coherency and enforce the phase decomposition  to \textit{RE}NiO$_3$  $\rightarrow$ \textit{RE}$O_x$+NiO~\cite{Gorbenko:2002p4026}. Moreover, as indicated by equation (1), the increase of film thickness ($h$) as well as  the larger lattice mismatch ($\epsilon$) may further  amplify the phase decomposition (i.e. $RE_2$O$_3$ and NiO) away from the single phase perovskite film~\cite{Novojilov:2000p2041}. 
Because the formation energy of \textit{RE}NiO$_3$ $[\Delta G_{RE\mathrm{NiO_3}} = \Delta G_{\mathrm{LaNiO_3}}+(h-sT) (r_{RE^{3+}} - r_{\mathrm{La}^{3+}}-\frac{1}{4}RT \ln(P)]$ also increases with  a decreasing tolerance factor~\cite{Chen:2015p031905}, the  epitaxial stabilization of the distorted members of the series becomes much more difficult. In recent years, several groups have overcome this challenge and  have successfully stabilized ultra-thin films of \textit{RE}NiO$_3$ with \textit{RE} = Pr, Nd, Sm and Eu~\cite{Ha:2012p233,Kareev:2011p114303,Feigl:2013p51,Liu:2010p233110,Meyers:2013p385303,Hauser:2015p092104,Jaramillo:2013p2455}.

 \subsection{Polarity mismatch:} 
 
The interest  in nickelate heterostructures also brought to light the issue of polar discontinuity at the subtrate-film interface that results in a diverging potential, which is    compensated  in  several ways , e.g. by electronic reconstruction,  cationic inter-mixing or  oxygen vacancies~\cite{Bristowe:2014p143201}.   In  a pure ionic model, because \textit{RE}NiO$_3$ consists of alternating [$RE$O]$^{1+}$, [NiO$_2$]$^{1-}$ atomic planes along the pseudo cubic (0 0 1) direction, this polarity issue was investigated  by growing a [1 uc LNO/1 uc LAO] superlattice  on non-polar STO ([SrO]$^0$, [TiO$_2$]$^0$) and polar LAO   ([LaO]$^{1+}$, [AlO$_2$]$^{1-}$)) (uc denotes pseudocubic unit cell).  Whereas the matched polarity on  the LAO substrate yields the desired +3 valence  of Ni, the polar mismatched case of LNO\ on  STO\ yields the first unit cell  with Ni in the  +2 oxidation state~\cite{Liu:2010p133111}.  Careful high-resolution transmission electron microscopy (HRTEM) combined with electron energy loss spectroscopy (EELS) confirmed Ni$^{2+}$O$^{2-}$ precipitation  driven by the interfacial polar fields~\cite{Detemple:2011p211903}.  This precipitation of the Ni$^{2+}$ containing layer  can be suppressed by the careful growth of  a buffer layer of LAO, which allows for  the preservation of Ni$^{3+}$~\cite{Liu:2010p133111} in all nickel containing layers. These experiments  highlight  the importance of  taking into  account the  polarity mismatch   in understanding material  properties in heterostructures.

\section{GROUND STATE ENGINEERING THROUGH HETEROEPITAXY}
Epitaxial stabilization   on such perovskite substrate not only provides the desired single crystalline materials, but also allows ground state engineering of nickelates by implementing the effect of strain, confinement, charge transfer etc.  The rapid progresses along these directions of research in recent times have been discussed in this section. 


\subsection{Epitaxial strain in ultra thin films}

Because epitaxial growth on a single-crystalline substrate forces the  film to have  the same in-plane lattice parameters during the atom-on-atom growth;  distortions,  lattice symmetry, octahedral tilts and rotations in the resulting thin film may become quite different from the bulk counterpart~\cite{Rondinelli:2012p261}. Epitaxial strain is conventionally quantified by $\epsilon=(a_{\mathrm{bulk}}-a_{\mathrm{sub}})/a_{\mathrm{bulk}}$, where $a_{\mathrm{bulk}}$ and $a_{\mathrm{sub}}$ are the pseudo cubic lattice constants of bulk $RE$NiO$_3$ and the substrate, respectively. However, as some chemical nonstochiometry may be introduced during the growth,  actual $\epsilon$ should be calculated using the experimentally measured lattice constant of an unstrained film (instead of $a_\mathrm{bulk}$). The effect of  strain on electronic and magnetic transitions of $RE$NiO$_3$ series has been thoroughly explored by diverse experimental probes including  thermal measurements and Hall effect measurement~\cite{Moon:2011p073037,Moon:2012p121106,Hauser:2015p092104}, synchrotron X-ray diffraction~\cite{May:2010p014110,Tung:2013p205112}, X-ray absorption spectroscopy~\cite{Chakhalian:2011p116805,Liu:2013p2714,Tung:2013p205112,Meyers:2013p075116,Bruno:2013p195108}, resonant soft X-ray scattering~\cite{Liu:2013p2714,Hepting:2014p227206,Catalano:2014p116110}, ARPES~\cite{Yoo:2015p8746,King:2014p443}, optical spectroscopy~\cite{Stewart:2011p176401,Stewart:2011p075125,Stewart:2012p205102,Hepting:2014p227206}.

\subsubsection{Orbital response to heteroepitaxy}

Magnetic interactions in transition metal oxides are determined by the specific geometry of chemical bonds,  the presence of various pathways for  exchange interactions, the orbital configuration of the transition metal ions, and $d$ electron occupancy. The strength and the nature (ferro- vs. anti-ferromagnetic) of these exchange interactions  can be summarized by the set of phenomenological rules, collectively known as Goodenough-Kanamori-Anderson rules, that reflect the strong intercoupling between spin, charge and orbital degrees of freedom. On this basis, the fundamental step to employing heteroepitaxial engineering is to alter  via the orbital-lattice interactions,  the  \textit{d}-band orbital  occupation and  energy level splittings caused by biaxial strain induced lattice deformations. The orbital  engineering in nickelates can be  rationalized in the following way:   coherent heteroepitaxy imposes a tetragonal distortion on the film,  thus removing the twofold $e_g$ orbital degeneracy due to the distortion of octahedral ligand field. Because both $d_{x^2-y^2}$ and $d_{3z^2-r^2}$ orbitals, constituting the degenerate $e_g$ orbitals in an octahedral crystal field, have even symmetry,    both tensile and compressive strains should symmetrically alter their energy positions relative to the strain-free energy band center. Although this
symmetric strain-induced orbital polarization (SIOP) ({\bf Figure~3(a)}) concept is a  common view for rationalizing orbital responses, surprisingly, ultra thin films of many perovskite oxides surprisingly do not follow this   simple notion~\cite{Tebano:2008p137401}. 


\begin{figure*}
\includegraphics[width=6.5in]{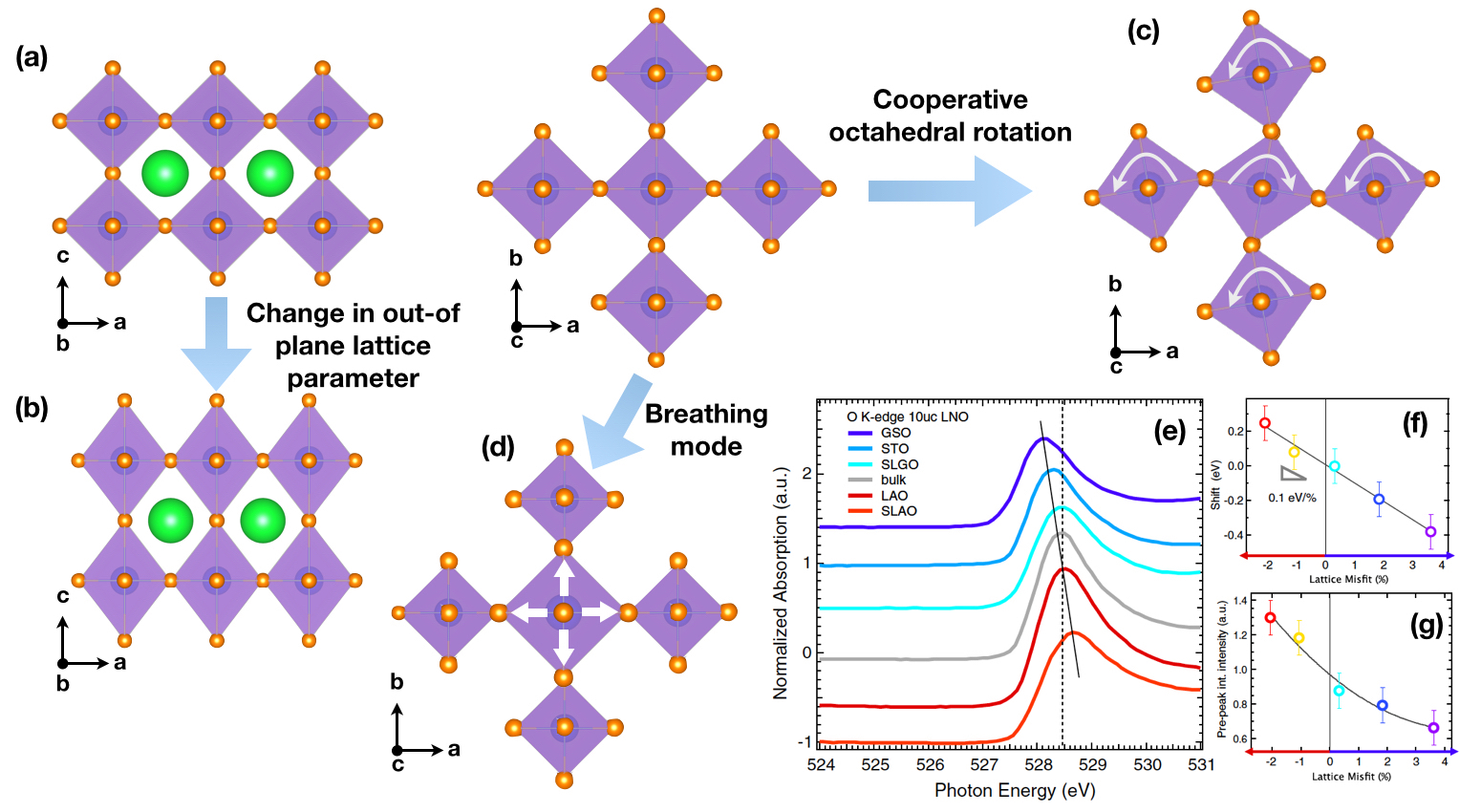}
\caption{(a)-(d) Different types of distortions for accommodating lattice
mismatch. (b) Out of plane expansion in an octahedron in accordance with  the concept of symmetric strain-induced orbital polarization. (c) Cooperative octahedral rotation. (d) Breathing mode. (e)-(g) Effect of strain-induced change in Ni-O covalency for LNO. (e)  Pre-peak in O $K$-edge absorption
for various strain states. Variation in (f) peak position and (g)  normalized
full width at half-maximum relative to  bulk LNO  with misfit $\epsilon$. Abbreviations: LAO, LaAlO$_3$;  STO, SrTiO$_3$; GSO, GdScO$_3$; SLGO, SrLaGaO$_4$; SLAO, SrLaAlO$_4$. Reprinted (panel (e)-(g)) with permission from Reference~\cite{Chakhalian:2011p116805} [J. Chakhalian et al., Phys. Rev. Lett. {\bf 107}, 116805 (2011).] Copyright (2011) by the American Physical Society.}
\label{lnostrain}
\end{figure*}

To  understand how the SIOP model is   violated, a 10 uc thick film of LaNiO$_3$ (LNO) was grown on LAO ($\epsilon$ = -1.1\%) and STO ($\epsilon$ = +1.8\%) substrates~\cite{Chakhalian:2011p116805}. For the LAO case, X-ray linear dichroism (XLD) measurement  showed a 100 meV lowering of the Ni $d_{3z^2-r^2}$  orbital compared to the $d_{x^2-y^2}$ orbital,  confirming the compression of  in-plane Ni-O bonds and  enlargement  of the out-of-plane (apical) Ni-O bonds, as expected from the SIOP model.  Following
the SIOP model, an inversion of  the $e_g$ orbitals was expected for tensile strain on STO. Surprisingly, no orbital polarization was detected for the STO case,  although   reciprocal space mapping  confirmed that the film was   fully coherent with in-plane lattice matched  to the STO substrate~\cite{Chakhalian:2011p116805,Tung:2013p205112}. This asymmetric orbital-lattice response to epitaxial strain signals  the existence of alternative compensation mechanisms for strain accommodation~\cite{Rondinelli:2012p261,PRB.90.045128}.    Density functional theory (DFT) calculations with the experimentally determined lattice constants showed the existence of single type NiO$_6$ octahedra  with different in-plane and out-of-plane Ni-O bonds  resulting in finite orbital polarization for  compressive strain. In contrast, tensile strain was preferentially  accommodated  by cooperative octahedral rotations  ({\bf Figure~3(b)}) and breathing mode distortions ({\bf Figure~3(c)}). The  importance of  these phonon modes is highlighted by the  observation of  breathing mode distortion, that causes a Ni-O bond-length disproportionation with alternating octahedra of Ni(1)O$_6$ (short Ni-O bonds) and Ni(2)O$_6$ (long Ni-O bonds), which  are all equivalent in the bulk LNO.  

 Further XLD experiments  on NNO films  revealed that contrary to LNO,  under tensile strain  the$d_{3z^2-r^2}$ orbital has a higher energy  compared to the $d_{x^2-y^2}$~\cite{Tung:2013p205112},  as expected from the  SIOP model for tensile strain. Such lifting of orbital degeneracy  was also seen by hard X-ray RIXS (resonant inelastic X-ray scattering) measurements, that identified a splitting of 0.4 eV between the $e_g$ states~\cite{Upton:2015p036401}.  This markedly different response of LNO vs. NNO   was attributed to the difference in octahedral rotational pattern of bulk: $a^-a^-a^-$ (LNO) and  $a^-a^-c^+$ (NNO) in Glazer notation, thus emphasizing the importance of lattice symmetry  and how it interacts with strain driven distortions of the structure~\cite{Tung:2013p205112}.

 \subsubsection{Strain induced self-doping}

Becasue \textit{RE}NiO$_3$  belong  to a part of $U$-$\Delta$ phase space with  an effectively negative charge transfer energy and a high degree of covalency, the ground state of $RE$NiO$_3$ is  a strong admixture of $d^7$ and $d^8\underline L$ electronic configurations ($\underline L$ is a ligand hole  on  the oxygen \textit{p }orbitals). This $d^8\underline L
$ state manifests itself as  a characteristic pre-peak around 528.5 eV in O $K$ edge (1$s$ $\rightarrow p$) X-ray absorption spectra (XAS).  Due to the weak core-hole interaction, the O $K$ edge XAS is an approximate measure of oxygen projected unoccupied density of states~\cite{Sarma:1996p1622}, and the intensity, position, and width of the pre-peak are the markers of  the degree of Ni-O bond covalency. As shown in {\bf Figure~3(e)}, for the series of LNO\ films subjected to  both tensile and compressive strain, the pre-peak revealed a systematic evolution in  the peak  energy position with the imposed amount of epitaxial strain. The direct strain control of ligand-hole density was deduced from  the nearly linear dependence of the peak position with $\epsilon, $  {\bf Figure~3(f)}.  The charge transfer energy
($\Delta$    (= e $\delta V_\mathrm{Mad}$ + I(O$^{2-}$) - A(M$^{\nu+}$) - $e^2$/d$_\mathrm{M-O}$ in the ionic model~\cite{Imada:1998p1039}, where  I(O$^{2-}$) is the ionization potential of oxygen,  A(M$^{\nu+}$) is the electron affinity of transition metal ion M$^{+\nu}$ ion and d$_\mathrm{M-O}$ is the nearest neighbor metal-oxygen distance)  depends on
the  relative Madelung  potential $\delta V_\mathrm{Mad}$ between Ni and O, so the movement of the prepeak highlights a decrease in  $\Delta$ with compressive strain. The corresponding decrease in  full width at half maximum (FWHM) with increasing tensile strain  {\bf Figure~3(g)} implies the narrowing of electron bandwidth ($W$). The strong   modulation in both $\Delta$ and $W$ implies a strain-induced unusual  self-doping effect, which was further substantiated by temperature-dependent Hall coefficient and thermoelectric power measurements~\cite{Moon:2012p121106}.
 
 \subsubsection{Strain driven phase engineering}  
As the nature of the ground states and  the transition temperatures depend strongly of Ni-O-Ni bond angle and Ni-O bond length, epitaxial strain  is  expected to  have a profound impact on the electronic and magnetic properties. To investigate this possibility, Liu et al.,~\cite{Liu:2010p233110,Liu:2013p2714} investigated   6 nm NNO films  as a function of epitaxial strain. The result shown in {\bf Figure~4(a)} indicates that unlike  the bulk where $T_{\mathrm{MIT}}=T_N$,  the transitions get separated, with the degree  of  separation scaling with   tensile strain. On the compressive side,  the temperature-induced MIT is completely quenched. Although several reports~\cite{Novojilov:2000p2041,Scherwitzl:2010fs,Caviglia:2012ea,Hauser:2013p182105} revealed the preservation  of  the MIT in compressively strained films, a  recent study by Hauser et al.~\cite{Hauser:2015p092104} demonstrated that  those films are actually partially relaxed, and optimally prepared films under compressive strain are entirely metallic~\cite{Hauser:2015p092104,Mikheev:2015xe}, consistent with the report of  Liu and coworkers~\cite{Liu:2010p233110,Liu:2013p2714}. These studies found the  resistivity of this anomalous metallic state  exhibits  non-Fermi liquid (NFL) behavior in the vicinity of a quantum critical point akin to the case of  bulk PrNiO$_3$ under external pressure~\cite{Zhou:2005p226602,Kobayashi:2015p195148}.  From those observations, it was concluded that film behavior under increasing  tensile strain resembles  the $A$-site replacement by a smaller $RE$ ion of the bulk, whereas the effect of compressive strain is very  similar to that  of  hydrostatic pressure. The exponent of NFL phase is also not affected by the change of film thickness~\cite{Mikheev:2015xe}. Although  resistance of these films surpasses the Mott-Ioffe-Regel limit at high temperature~\cite{Jaramillo:2014p304}, unlike the case for  the cuprates $\rho$ eventually  saturates for nickelates. This behavior  was shown to related with the orbital splitting between $d_{x^2-y^2}$ and $d_{3z^2-r^2}$ orbitals~\cite{Mikheev:2015xe}. . The effect of epitaxial strain was also investigated for the distorted  \textit{RE}NiO$_3$ members with $T_{\mathrm{MIT}} > T_N$~\cite{Meyers:2013p075116,Bruno:2013p195108,Catalano:2014p116110}.  Similar to  the  NNO case, the separation between $T_{\mathrm{MIT}}$ and $T_N$~ decreases as the amount of tensile strain is lowered and  the films grown under large compressive strain becomes entirely metallic~\cite{Catalano:2014p116110}.   

\begin{figure*}
\includegraphics[width=6.5in]{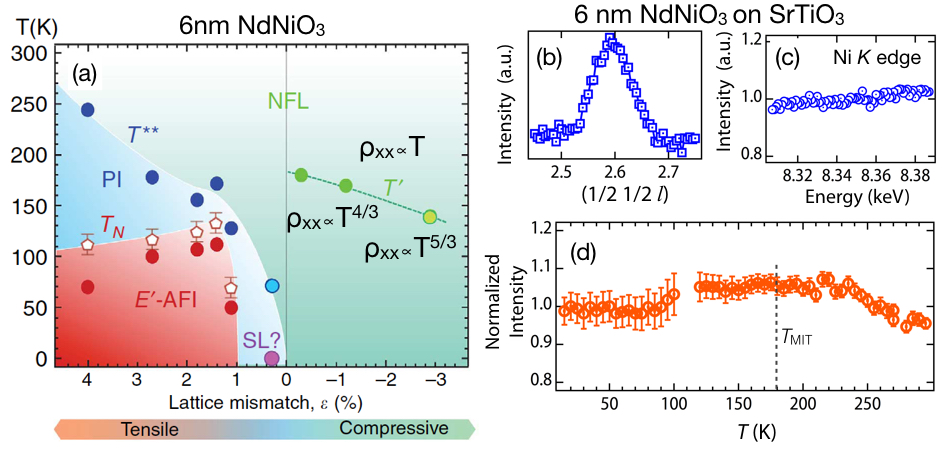}
\caption{(a) Lattice mismatch-temperature phase diagram for 6 nm thin NNO film. Panel (a) adapted  from Reference~\cite{Liu:2013p2714} NFL, PI and AFI denote non-Fermi liquid, paramagnetic insulator and AFM insulator, respectively. {\it T**} (blue), $T_0$ (green) and the hysteric inflection point (red) are denoted by closed circles, whereas $T_N$ is denoted by open pentagon and its error bar is defined as the temperature step size, 10 K.   (b) $l$ scan through (1/2 1/2 5/2)$_\mathrm{pc}$ reflection for NNO on STO substrate. Due to the twinned domain structure of the film, this reflection is a combination of (105)$_\mathrm{ortho}$, (015)$_\mathrm{ortho}$, (231)$_\mathrm{ortho}$, (321)$_\mathrm{ortho}$ reflections. (c) Ni does not contribute to this reflection even in the insulating phase as demonstrated by the energy scan around the Ni $K$-edge. (d) Temperature dependence of this reflection, recorded during the heating run from low temperature. For details of panels (b)-(d), see Reference ~\cite{Upton:2015p036401}.  The absence of Ni contribution to the ($h$ 0 $l$)$_{\mathrm{ortho}}$   reflection with odd $h$ and $l$ in such ultrathin films was also  confirmed for single-domain NNO film, grown on a NdGaO$_3$ substrate~\cite{Meyers:2015p07451}.  }
\label{nnostrain}
\end{figure*}

\subsubsection{Selective suppression of order parameters through epitaxy}
Although both the antiferromagnetic order and the MIT are  preserved in epitaxially grown  NNO films with $\epsilon >+ $1.1\%, the next important question is whether the proximity to  the substrate of different  crystal  symmetry can alter the orthorhombic to monoclinic symmetry lowering across the MIT   and the concomitant  charge ordering. The charge ordering scenario in the bulk-like  NNO\ films  was investigated with  resonant X-ray scattering experiments at the Ni \textit{K}-edge~\cite{Staub:2002p126402}.  In the high-symmetry orthorhombic metallic phase Ni sites do not contribute to particular reflections, which gain a Ni contribution in the low temperature monoclinic phase and can be used to study the details of the charge ordering phase via the resonant signature~\cite{Staub:2002p126402,Lorenzo:2005p045128,Scagnoli:2005p155111}.   An interesting recent result is that, for an ultrathin (6 nm) NdNiO$_3$ film grown under tensile strain, even though the film displayed an MIT and magnetic ordering, there was no resonant Ni signal for these reflections even in the insulating phase ({\bf Figure~4(c)}). This observation, together with the absence of any detectable temperature dependence of the reflection ({\bf Figure~4(d)}), led to the conclusion   that $Pbnm$ symmetry is preserved deep  into the insulating state~\cite{Upton:2015p036401,Meyers:2015p07451}.  This  result implies the exceptionally rare case of a purely  electronic Mott-Hubbard MIT that is driven by the magnetic ordering. 

Despite the absence of   both the  lattice symmetry change and Ni charge disproportionation,   the transition was found to be accompanied by  a peculiar  charge redistribution from the Ni 3$d$ to Nd 5$d$ orbitals that may play a role in the MIT~\cite{Upton:2015p036401}. 
Although the relevance of Mott-Hubbard picture about the nature of the MIT was highlighted by  the mid-gap spectral weight transfer  in optics measurements~\cite{Stewart:2011p176401}, the charge redistribution which involves the \textit{A}-site states  questions the  simple picture of the conventional Mott  transition. The $K$-edge RIXS measurement  and  the resonant x-ray scattering  data suggest   that the MIT\ is neither   pure Mott-Hubbard nor charge-transfer.

In contrast to the case for NdNiO$_3$,  the bulk-like charge ordering pattern remains unaltered in ultra-thin films of highly  distorted EuNiO$_3$~\cite{Meyers:2015}. Those experiments  stress    the exquisite sensitivity of the electronic  structure in the insulating state  to various type of structural distortions ~\cite{Balachandran:2013p054101} including the Jahn-Teller distortion (NdNiO$_3$) and  breathing mode distortion (EuNiO$_3$),  under epitaxial constraint. The issue of  charge ordering in such geometry  was also investigated by Raman scattering~\cite{Hepting:2014p227206}, and  phonon modes characteristic of the charge ordering were indeed observed in insulating state of thin film and superlattices of PrNiO$_3$ (of thickness $\sim$ 12 nm)  under tensile strain.  Interestingly, these Raman modes were absent in  superlattices under compressive strain, which are weakly metallic but retain   $E'$ antiferromagnetic ordering. Such an antiferromagnetic, metallic phase does not exist in the bulk nickelate and  has been referred to as a spin-density wave phase~\cite{Hepting:2014p227206}, which was predicted theoretically   by  Lee and co-workers~\cite{lee-prl,Lee:11}.

\subsection{Quantum confinement in heterostructures}

The ability to grow these materials  with sub-unit cell precision  has opened  another new dimension  to explore highly two-dimensional electronic and magnetic phases in the quantum confinement regime. The SL structure consisting of an electronically active material and a wide band gap insulator  (one side can be vacuum) blocks electron hopping  along the confining direction (i.e. across the interface), the specific choice of confining interface further  determines the boundary condition in this  quantum well geometry. In this section, we explore how heterostructures can be used to tune the electronic and magnetic states of nickelates.

\subsubsection{Dimensionality and Carrier Localization}

The correlated metallic behavior of bulk LNO  provides a perfect play ground to test the itinerant behavior of \textit{d}-electrons down to  the  atomic scale. Toward this goal, several group explored the effect of  dimensionality along  with  electron-electron correlations was explored by growing a series of single layer films and SL structures of LNO of different thicknesses~\cite{Son:2010p062114,Scherwitzl:2011p246403,Liu:2011p161102,Boris:2011p937,Sakai:2013p075132,Kumah:2014p1935,King:2014p443}. As an example, consider the results in {\bf Figure~5(a)}, showing the temperature dependent resistivity behavior for a series of LNO films grown on LAO substrates by oxide molecular beam epitaxy~\cite{King:2014p443}.  Although the room temperature conductivity  decreases with decreasing film thickness, the metallicity over the entire temperature range is  maintained  down to 4 uc. Upon approaching the 2 uc limit, the LNO film becomes insulating. Although  all  groups have reported this confinement induced MIT, the critical film thickness at which the insulating behavior sets in is  strongly  dependent  on  the strain value and the architecture (i.e. film vs. SL). With respect to dimensionality effects in single layer films, one issue is the presence (or lack) of a polar discontinuity, which may explain the differences between the results on LAO~\cite{King:2014p443} \textit{vs.} STO~\cite{Scherwitzl:2011p246403}. Resonant X-ray absorption spectroscopic measurements on Ni $L_{3,2}$-edge complemented by first  principle calculation~\cite{Liu:2011p161102} and soft, hard X-ray photoemission spectroscopic measurements~\cite{Gray:2011p075104,Sakai:2013p075132} revealed a gradual evolution of electronic structure including the charge transfer gap. X-ray diffraction showed a change in atomic arrangement,  in terms of both decreased Ni-O-Ni bond angle and lattice rumpling near the interface with a reduced number of  unit  cells~\cite{Kumah:2014p1935}. In addition, layer-resolved studies established that the suppression of DOS is   more pronounced near the interface~\cite{Kaiser:2011p116402}, where the octahedral tilts are larger than in layers away from the interface~\cite{Hwang:2013p060101,Kumah:2014p1935}. All of these results indicate at the bandwidth control is an important component for the MIT as is the case for bulk nickelates~\cite{Medarde:1997p1679,Catalan:2008p729}.
   
   \begin{figure*} 
\includegraphics[width=6.5in]{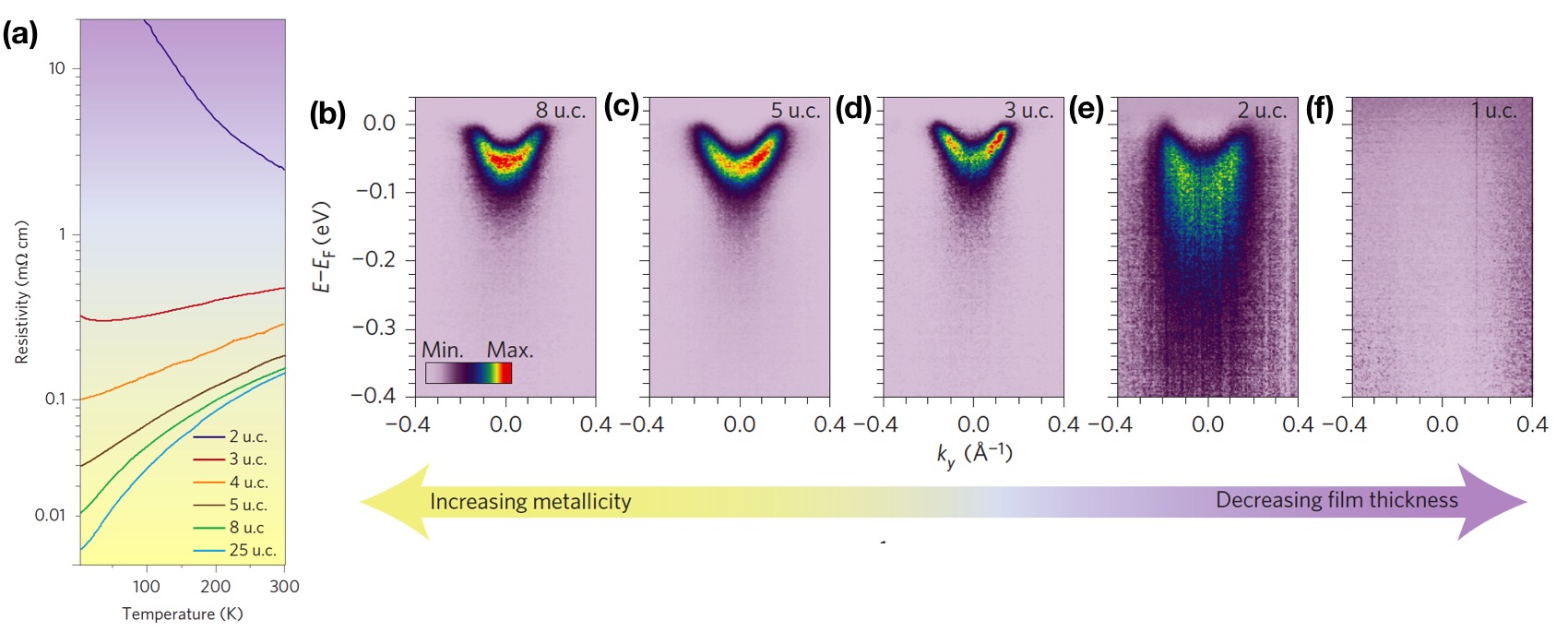}
\caption{ (a) MITs as a function of LNO layer thickness. (b-f) The evolution of electronic structures across MIT is probed by (b-f) angle-resolved  spectroscopy along (0.5$\pi/a$,$k_y$,0.7$\pi/a$).  Abbreviations: LAO, LaAlO$_3$; LNO, LaNiO$_3$; MIT, metal-insulator transition.
Reprinted by permission from Macmillan Publishers Ltd: [Nature Nanotechnology] (reference~\cite{King:2014p443}), copyright (2014).}
\label{confinement}
\end{figure*}

Recent  ARPES   studies~\cite{King:2014p443}  revealed the retention of bulk-like electronic structure and Fermi liquid characteristics (see narrow parabolic band in ({\bf Figure~5(b)-(d)}) down to 3 u.c. thin film and the thickness dependent MIT was attributed  to the presence of a spin/charge instability in two dimension~\cite{King:2014p443}.

\subsubsection{Magnetism in Confined Architectures}

Because all bulk rare earth nickelates  that display a MIT also undergo magnetic
transition, the observation of the insulating state in the few unit cell limit  prompted  several  groups to investigate  its   spin degree of freedom.   Ultrathin superlattices were initially investigated by muon spin rotation ($\mu$SR) ~\cite{Boris:2011p937}, and metallic  4-uc LNO/4-uc LAO SLs were found to remain  paramagnetic down to the lowest temperatures, as for the case of  bulk LNO, but  antiferromagnetism was clearly detected  in the insulating phase of a 2-uc LNO/2-uc LAO SL. Because $\mu$SR cannot determine the ordering pattern, a series of LNO SLs with dielectric spacers were grown on several substrates and were investigated  by resonant soft x-ray diffraction\cite{Frano:2013p106804}. {\bf Figure~6(a)} shows the resulting \textit{L}-scans around  (1/4 1/4 1/4)$_{\mathrm{pc}}$ (where the subscripted pc denotes pseudocubic)  [(1/2 0 1/2) in the orthorhombic setting] with the incident photon energy tuned to the Ni $L_3$ edge for  $N$ uc LNO/$N$ uc $AB$O SLs with $N$= 2, 3, 4 and $AB$O = LAO,  DyScO$_3$. The experiment clearly  established  the presence of bulk-like $E'$-type antiferromagnetic order for all of the  SLs with $N$=2. The strong dependence of spin direction on underlying
structure   obtained from the analysis of polarization dependence and azimuthal scans,   ({\bf Figure~6(c), (d)}) further emphasized the importance of spin-phonon coupling. Because this non-collinear  antiferromagnetic structure has the same propagation wave vector as the Fermi surface nesting vector obtained from calculations and  the high temperature phase  is metallic ({\bf Figure~6(b)}), this magnetic state has been interpreted as spin-density   wave. The disappearance of the spin-density  wave states for $N$ = 3, 4 SL was attributed  to the increasing three-dimensional Fermi surface that  in turn suppresses the tendency  for nesting together with the onset of the metallic phase. Such a nesting driven mechanism behind magnetic ordering was further corroborated  by very recent ARPES measurements~\cite{Berner:2015p125130}. Similar polarization-dependent magnetic scattering experiments with azimuthal scans for thin films spanning the whole $RE$NiO$_3$ series  are clearly necessary to probe  any  variation in the spin orientation with the change in the Ni-O-Ni bond angles.                                                                         

\begin{figure*}
\includegraphics[width=5.in]{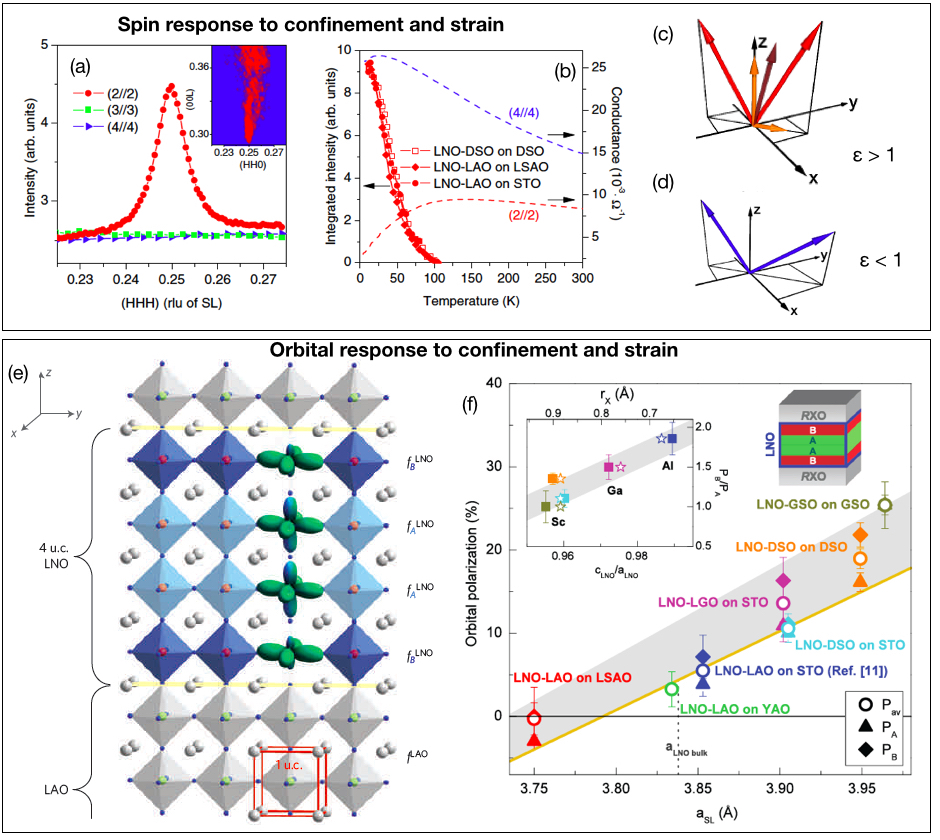}
\caption{(a)-(d) Effect of confinement and epitaxial strain on Ni spins,   reprinted  with permission from Reference~\cite{Frano:2013p106804} [A. Frano et al. Phys. Rev. Lett. {\bf 111}, 106804 (2013).] Copyright (2013) by the American Physical Society. (a) $L$ scans around  around {\bf q$_0$} =(1/4 1/4 1/4)$_{pc}$ at 10 K and $E$ = 853.4 eV for $N$ u.c. LNO/$N$ u.c. $AB$O SLs  ($AB$O= LAO, DSO).  Reciprocal-space map of the scattered intensity from the $N$ = 2 SL. (b)  Temperature dependence of the magnetic Bragg intensity at  {\bf q$_0$} for  LNO-based SLs with $N$ = 2 emphasizes a paramagnetic to AFM transition.  dc electrical conductance measurements  for $N$ = 2 and $N$ = 4 SLs are also shown. (c) Spin directions  of $N$=2 SL under tensile strain. Red arrows (equal moments on each site are symmetrically tilted from [001] axis by 28$\pm$2$^{\circ}$) and orange arrows (two spin sublattices: a large moment along [001] and a small moment along [110]) show two possible arrangements. (d) The magnetic moments for $N$=2 SL under compressive strain is similar to the those of the bulk~\cite{Scagnoli:2006p100409}. (e) Schematic structure to highlight different orbital configurations of the interfacial layer compared with the orbital configurations of the bulk-like layers, reprinted by permission from Macmillan Publishers Ltd: [Nature Materials] (reference ~\cite{Benckiser:2011p189}) copyright (2011). (f) Orbital polarization as a function of substrate lattice parameter for various LNO-based superlattices. Abbreviations: LAO, LaAlO$_3$; LNO, LaNiO$_3$; pc, pseudocubic; SL, superlattice; STO, SrTiO$_3$; DSO, DyScO$_3$; LSAO, LaSrAlO$_4$, GSO, GdScO$_3$; LGO, LaGaO$_3$; YAO, YAlO$_3$. Panel (f)  reprinted  with permission from Reference~\cite{Wu:2013p125124} [M. Wu et al., Phys. Rev. B {\bf 88}, 125124 (2013).] Copyright (2013) by the American Physical Society.
 }

\label{orbital}
\end{figure*}

\subsubsection{Orbital polarization}

From the electronic structure standpoint, due to the strong similarity between $RE$NiO$_3$ and high -$T_c$ cuprates (i.e. one electron  $vs$. one hole on the $e_g$orbital), Chaloupka \&  Khaliullin~\cite{Chaloupka:2008p016404} proposed, as an   analog for high $T_c$ superconductors, a single unit cell LNO layer heterostructured with a band insulator.  In their model, it was assumed that  tensile strain  would enforce the   electron   in  to the   $d_{x^2-y^2}$ state and the insulating barrier across the interface reduce  the bandwidth of the $d_{3z^2-r^2}$ band by blocking the out-of plane hopping. More recent  LDA (local density approximation) + DMFT calculations corroborated the possibility of a  cuprate like Fermi surface  in such heterojunction~\cite{Hansmann:2009p016401}.

This  theory proposal   sparked    a vivid interest in  orbital physics of $RE$NiO$_3$ compounds. Apart from  strain and quantum confinement, the choice of non transition metal ion $X$ in the LNO/La$X$O$_3$ heterostructure was  shown be an important factor in controling the  degree of  orbital polarization~\cite{Han:2010p134408}. Recall that, in the bulk  ligand holes are equally distributed over the six-oxygen  surrounding a central Ni ion.   In the LNO/LaXO$_3$ SL  because of the strong  ionic nature of \textit{X } the holes initially  located  on the apical oxygen (i.e. Ni-O-$X$ bond) would be redistributed  back into  the $ab$ plane. This redistribution of holes  would  be akin  to  \textit{d}-band filling having  a strong effect on the \textit{d}-orbital occupancy, altering the magnitude  and even switching the sign of  orbital polarization \cite{Han:2010p134408}.  After the charge transfer physics between Ni \textit{d-} states and O \textit{p} states was explicitly incorporated into  DFT+DMFT calculations, the resulting orbital polarization became considerably  reduced~\cite{Han:2011p206804}.
 

The experimental work on orbital response to confinement and epitaxial strain,   revealed  the behavior that is  significantly more complex than  anticipated from the theory. For example, 1-uc LNO/1-uc LAO SLs grown under compressive strain on LAO showed equal electronic population of   $d_{x^2-y^2}$ and $d_{3z^2-r^2}$ orbitals  even though the orbitals an energy splitting of $\sim 100$ meV~\cite{Freeland:2011p57004}.  Disa et al\cite{Disa:2015p026801} also recently confirmed this zero orbital polarization. In contrast, the same SL grown under tensile strain on STO showed only $\sim5$\% orbital polarization and most importantly no splitting between the $e_g$ orbitals.  \textit{Ab-initio}
calculations~\cite{Freeland:2011p57004,Blanca:2011p195450} found that the coordination between the Al and the Ni resulted in a chemical mismatch that influenced the hole density on the apical oxygen and subsequently the Ni $d_{3z^2-r^2}$ orbital. A new approach termed orbital reflectometry (a combined analysis of XLD and resonant X-ray reflectivity)~\cite{Benckiser:2011p189} applied to    a 4-uc LNO/4-uc LAO   superlattice (SL) on a STO substrate also detected a similar value of the orbital polarization. As illustrated in {\bf Figure~6(c)}, the power of this method is the ability  to differentiate the orbital polarization across the layers. By careful choice of spacer layers and application of high tensile strain, large orbital polarization (up to 25\%) was reported for a 4-uc LNO/4-uc GdScO$_3$  SL~\cite{Wu:2013p125124}  ({\bf Figure~6(d)}). A very recent  study of a  4-uc PrNiO$_3$/4-uc PrAlO$_3$ SL revealed  the presence of  smaller orbital polarization ($P_\mathrm{av}$) of $\sim$ 5\%~\cite{Wu:2015p195130}, which also decreases across the MIT. Interestingly, no significant temperature dependence of $P_\mathrm{av}$  was observed across the magnetic transition of  compressively strained SLs, which had earlier been assigned as a spin density wave phase~\cite{Hepting:2014p227206}.

 \subsection{Interfacial charge transfer}
The redistribution of charge at the interface between two dissimilar materials is one of the key concepts behind modern electronics. The  difference in the chemical  potential may cause  charge transfer across  the interface;  which in-turn it can alter  spin and orbital sectors, resulting in novel electronic and magnetic states emerging in complex oxide heterostructures~\cite{Hwang:2012p103,Chakhalian:2012p92,Chakhalian:2014p1189,Bhattacharya:2014p65}. For example, electronic reconstructions have been explored for LNO  by forming  interface with manganites~\cite{May:2009p115127,Gibert:2012p195,Hoffman:2013p144411,Grutter:2013p087202,Di:2015p156801,Piamonteze:2015p014426} and titanates~\cite{Chen:2013p116403,Disa:2015p026801,Cao:2015}. 

Although bulk CaMnO$_3$ (CMO)\ is an antiferromagnetic insulator, interfacial ferromagnetism limited within one unit cell of CMO from the interface was uncovered in $N$-uc CMO/$M$-uc LNO  superlattices  when  the thickness ($M$ $\ge4$)    was   sufficient to make the system metallic~\cite{Grutter:2013p087202}. Because no sizable charge transfer was observed, the origin of ferromagnetism connected to the appearance of  metallicity was attributed to double exchange between Mn$^{4+}$ and Ni$^{3+}$ along the interface.  A ferromagnetic coupling between the interfacial layers was also explored in 2-uc LaMn$^{3+}$O$_3$/2 uc LaNi$^{3+}$O$_3$] SLs, in which an electron transfer  from Mn to Ni led to the formation of a Mn$^{4+}$ -  Ni$^{+2}$ pair~\cite{Hoffman:2013p144411}.  In addition, a helical magnetic state without any spin-orbit coupling or Dzyaloshinsky-Moriya
interaction,  emerging from this interfacial charge transfer~ in LNO layers have been recently reported~\cite{Hoffman:2014}. When prepared along the [111] direction, LNO/LaMnO$_3$ SLs were also   displayed exchange bias, which was not seen for the  [001]  oriented case~\cite{Gibert:2012p195}.

As a way to engineer electronic correlations, the interface between Mott insulator LaTi$^{3+}$O$_3$ and correlated metal LNO was  recently proposed~\cite{Chen:2013p116403}.  The follow-up  experiments confirmed that such a heterojunction exhibits  massive  interfacial charge transfer from Ti to Ni sites resulting in an overall insulating ground state with a charge  excitation gap of 0.2 eV  between Ni $d$ and Ti $d$ states. The correlated gap between  UHB and LHB for Ni was  determined to be 1.5 eV~\cite{Chen:2013p116403, Cao:2015}.  After  inserting extra LaTiO$_3$ layers into  a single unit cell LNO/LAO heterostructure, an unusually high degree of orbital polarization ($\sim$50\%)  on Ni sites has been  reported and connected to the charge transfer between Ti and Ni~\cite{Chen:2013p186402,Disa:2015p026801}.

\subsection{Geometrical Engineering}

All  films and heterostructures  described above were synthesized along the pseudocubic   [001]  direction. Whereas  along the [001]$_{p.c.}$ direction $AB$O$_3$  consists of alternating $A$O, $B$O$_2$ atomic planes, the same perovskite  along the [111]$_{\mathrm{pc}}$ direction consists of alternating [$A$O$_3$] and $B$ planes. Thus, by growing two  pseudocubic  unit cells of $AB$O$_3$  along  [111]$_{\mathrm{pc}}$  one can generate a  new lattice with  two vertically shifted triangular planes of $B$ sites ({\bf Figure ~7(a)}).  As illustrated in \textbf{Figure ~7(b)}, this artificially generated buckled honeycomb   lattice provides a  unique opportunity to investigate  physics of $d$ electrons in  graphene-like geometry~\cite{Xiao:2011p596,Ruegg:2011p201103,Yang:2011p201104}. Similarly, the dice lattice shown in {\bf Figure ~7(c)},   can be obtained by growing  three  unit cells~along the [1 1 1]$_{\mathrm{pc }}$ direction\cite{Wang:2011p241103}.  

Due to the   difference in periodic arrangements of atoms along  [111]  vs. [001], the theoretical calculations ~\cite{Ruegg:2011p201103,Yang:2011p201104,Ruegg:2012p245131,Ruegg:2013p115146,Doennig:2014p121110} for (111)-oriented bilayers of nickelates have  predicted  the realization of several exotic phases unattainable in either  bulk or  (001)-oriented heterojunctions. For example, the model Hamiltonian calculations in the strongly correlated limit predicted that orbital ordering wins spontaneously over the bulk-like charge ordered phase ~\cite{Ruegg:2011p201103,Ruegg:2012p245131,Ruegg:2013p115146,Doennig:2014p121110},  whereas   in the weakly correlated limit
 a number of  topological phases ($e.g.$ Dirac half-metal phase, quantum anomalous Hall insulator phase, and spin nematic phase)  driven exclusively    by interactions and without    spin-orbit coupling were predicted~\cite{Ruegg:2011p201103,Yang:2011p201104}.

\begin{figure*}
\includegraphics[width=6.0in]{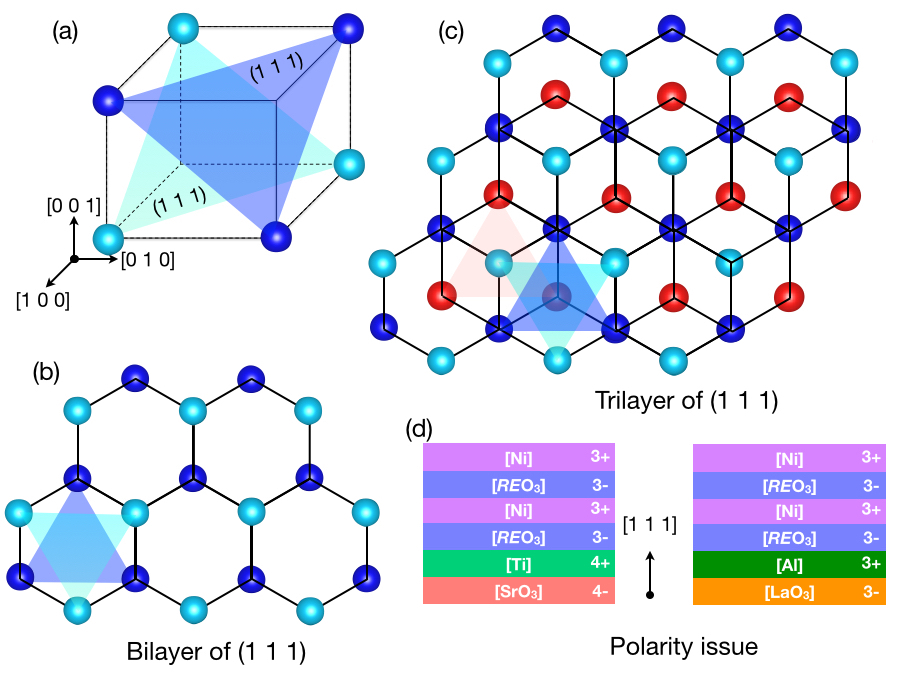}
\caption{(a) (111) planes of $B$-sites for $AB$O$_3$ perovskite structure. (b,c) Bilayer (b) and trilayer (c)  of such (111) planes generate a buckled honeycomb and dice lattice, respectively. For visual clarity, $A$ and oxygen sites are not shown. (d) Schematics of polar discontinuity (in an ionic picture and without considering any surface reconstruction) for the growth of $RE$NiO$_3$ on STO and LAO  (111) substrates. Abbreviations: LAO, LaAlO$_3$; STO, SrTiO$_3$. }
\label{geometry}
\end{figure*}

Because all  popular perovskite substrate such as STO (LAO, NdGaO$_3$, YAlO$_3$)  are  polar    along [111] with alternating +4$e$,  -4$e$ (+3$e$, -3$e$) charges per plane,   epitaxial thin film growth along this direction is rather undeveloped due to the possibility  of complex structural, chemical and electronic reconstructions that   act to compensate for the polarity jump. To elucidate the issue,  the unit cell by  unit cell growth was investigated by monitoring the growth progression with in-situ reflection high energy electron diffraction (RHEED) of  LNO on STO  (polar interface) and LAO (non-polar interface)  (111) substrates~\cite{Middey:2014p6819}. The measurements  showed the presence of non-perovskite phases within the  first 5 unit cells on the STO substrate; a combination of X-ray diffraction and  resonant X-ray spectroscopy identified the formation of  La$_2$Ni$_2$O$_5$ phase. In contrast, a stoichiometric LNO was successfully grown on a LAO (111) substrate, thus confirming that the absence of the polar mismatch at the interface is the key for the   [111] oriented heterostructures~\cite{Middey:2012p261602}.

Followed by the theoretical  predictions~\cite{Ruegg:2011p201103,Yang:2011p201104,Ruegg:2012p245131,Ruegg:2013p115146,Doennig:2014p121110}, the behavior of $d^7$ Mott electrons  on the honeycomb lattice   was investigated by growing 2-uc NNO/4-uc LAO  SL on a LAO (111) substrate. The resultant system exhibits a new ground state characterized by  antiferromagnetic correlations and  orbital ordering, and this lattice-geometry stabilized new ground state is unattainable in either bulk NNO or in analogous heterostructures grown along the conventional [001] direction~\cite{Middey:2014p1570}.  Although design of  \textit{RE} nickelates with  new lattice geometries  is still in its  infancy, the potential of this interesting concept   for the ground state engineering was recently highlighted also by a report of  markedly  larger charge transfer in [\textit{n} LNO/\textit{n} LMO] (111) SLs with \textit{n} = 5, 7  compared to the equivalent  (001) SLs~\cite{Piamonteze:2015p014426}.

 \section{THEORETICAL UNDERSTANDING}

The most important task of the theory of the $RE$ nickelates  is to link the atomic physics of the strongly correlated Ni $d$-orbitals to the behavior of actual materials. Issues of current importance include the nature of the MIT and its connection to lattice distortions and to magnetic ordering, the degree to which orbital occupancy can be varied by strain or other external parameters, and  the nature and origin of the observed magnetic states.  

An important issue in the context of transition metal oxides generally and  $RE$NiO$_3$ materials in particular is  orbital engineering: the degree to which the electronic structure  and, in particular the occupancy of different $d$-orbitals and  the number of bands crossing the Fermi level  can be changed by strain or by heterostructuring \cite{Chakhalian:2014p1189}. Following the  proposal of Chaloupka and Khalliulin \cite{Chaloupka:2008p016404} that under appropriate circumstances LNO might exhibit a one-band Fermi surface analogous to that of the high-$T_c$ cuprates, many groups~\cite{Wu:2013p125124,apl.1.4881557, Wu:2015p195130,Chakhalian:2011p116805,Tung:2013p205112,Freeland:2011p57004,Benckiser:2011p189} have investigated the degree to which orbital occupancies of $RE$NiO$_3$ materials can be manipulated. As far as is known, in bulk materials the relevant Ni $d$-orbitals are roughly equally occupied. The key finding~\cite{Chakhalian:2011p116805,Tung:2013p205112,Freeland:2011p57004,Benckiser:2011p189,Wu:2013p125124,Wu:2015p195130} is that in strained SLs a differential orbital occupancy of up to at most 25\% can be achieved. Understanding the limits of orbital engineering  is a second challenge to theory. 

We begin the theoretical discussion by consideration of the electronic configuration of  Ni sites, which  is a superposition of  states with 6, 7, 8, or 9 Ni $d$-electrons. We can schematically write
\begin{equation}
\left|\psi_{Ni}\right>=\alpha\left|d^6\right>+\beta\left|d^7\right>+\gamma\left|d^8\right>+\delta\left|d^9\right>
\label{Niconfig}
\end{equation}
The Hund's and ligand field energies are such that the $t_{2g}$-symmetry $d$ orbitals are fully filled and the relevant electronically active Ni $d$ orbitals are the $e_g$ states. Thus $d^6$ corresponds to an empty Ni $e_g$ shell, $d^7$ to an $e_g$ shell with one electron, and so forth. Although the Hund's coupling is evidently not strong enough to overcome the ligand field, it is believed to be strong enough to ensure that the dominant $d^8$ configuration is high-spin (\textit{S}=1, with one electron in each $e_g$ orbital).

The nickelates are believed to be   strongly correlated  materials in which interaction effects beyond  DFT  play a key role in the physics. As a simple example, in DFT the occupation probabilities of the different Ni valences (squares of the coefficients in Equation~\ref{Niconfig}) are determined by the mean occupancy and non-interacting electron statistics; in the actual materials the mean occupancy may be different from that predicted by DFT, and the relative probabilities of different states are likely to be strongly affected by interactions.   However, results of DFT calculations and their  +$U$  extensions   play an important role in our understanding of the materials and, at this stage, provide the only convenient mechanism for comprehensive structural relaxation, and are the starting point for   beyond DFT many-body calculations.  

The basic picture emerging from standard DFT calculations is that the mean Ni $d$-valence   slightly greater than $8$ \cite{Han:2011p206804} and that in the ideal structure two bands, of mixed Ni $e_g$ and $O_{2p}$ origin, cross the Fermi level (in the actual materials, monoclinic or GdFeO$_3$-type distortions increase the unit cell so that it contains 4 Ni ions; the associated backfolding leads to multiple band crossings). Structural relaxation calculations indicate an increasing degree of GdFeO$_3$ distortion as the $RE$ ion is changed from Nd across the series to Lu, consistent with experiment, and provide valuable information regarding rotational changes  in SLs~\cite{Balachandran:2013p054101}.  Pure DFT calculations generally predict a metallic and non-magnetic ground state in all members of the family. 

The  +$U$  extensions of DFT do find a charge ordered ground state for reasonable values of $U$~\cite{Mazin:07} but greatly overestimate the critical pressure required to destroy the charge ordering.\cite{PRB.89.245133,PRB.90.235103}. Jahn-Teller effects associated with lifting the degeneracy of the two $e_g$ states were considered by He \& Millis  \cite{He15} and were found to become relevant at large ($\sim 4\%$) strains. In addition, the +$U$ methodologies predict a ferromagnetic ground state, rather than the long-period $E$ or $E^\prime$ state observed experimentally.  However, for a reasonable range of $U$, Bellaiche and co-workers  \cite{bellaiche} were able to stabilize the $E'$ antiferromagnetic state (two inequivalent Ni atoms, one, associated with the long-bond octahedron and  carrying a non-zero   moment and, the other associated with the short-bond octahedron and  carrying  zero moment) as a metastable state.    Puggioni {\it et al.} \cite{Puggioni12} as well as Bruno {\it et al.} \cite{Bruno:2013p195108}  used hybrid functional methods  for strained films of the nickelates to show that the experimentally observed ground state is reproduced in these calculations.  

An alternative family of theoretical approaches uses model calculations based on simple physical pictures following  the assumption that, in the ground stat,e only one   valence  is dominant. The   Mott-Hubbard picture is based on the assumption that the relevant valence is $d^7$.  
In this picture, the main charge transfer process involves removing an electron from one site (creating a $d^6$ configuration) and adding it  to an adjacent site, creating a $d^8$ configuration, so that  the key energy is $U=E(d^8)+E(d^6)-2E(d^7)$. One expects $U>0$, in which case the important physical effects are the Mott transition (conductivity is blocked if $U$ is too large relative to the bandwidth)  and Jahn-Teller physics associated with cubic-tetragonal distortions that  lift the degeneracy of the $d^7$ state.  Also important is the Hunds coupling, $J_H$, which favors high-spin configurations. 

To study  this situation one considers a Hubbard-like model, with two bands (representing the Ni $e_g$ states, or, in a more sophisticated interpretation, the $e_g$-O$_{2p}$ antibonding bands, and that are subject to  the $U$,   $J_H$ and perhaps an electron-phonon interaction. Several researchers have taken this approach~\cite{Mazin:07,Chaloupka:2008p016404,lee-prl,Lee:11,Hansmann:2009p016401}.  Mazin and co-workers \cite{Mazin:07} (using an approach based much more closely on DFT calculations) observed  that at the crossover between itinerant and localized behavior ( with a bandwidth comparable to the  Coulomb repulsion) interesting physics can occur; in particular,  if the intra-atomic exchange interaction strength $J_H$ is large enough, it may drive  charge ordering. 

Hansmann and collaborators \cite{Hansmann:2009p016401} studied the Mott-Hubbard  picture, using a DFT+DMFT methodology in which the correlated orbitals were  built from  the near Fermi surface $e_g$-derived conduction bands. They found that for the generally accepted values of $U$ and $J$ the materials were near a Mott MIT. No indication of charge ordering was found. A superlattice-induced breaking of the cubic ($O_h$) point symmetry down to tetrahedral ($T_{d}$) was found to drive a very substantial orbital polarization, leading, for reasonable parameters, to a single-band Fermi surface. 

Lee and co-workers \cite{lee-prl,Lee:11} addressed the weak-coupling physics of orbitally degenerate Mott-Hubbard systems by using a Hartree-Fock analysis of a two-band model with a band structure corresponding to  the calculated near Fermi surface bands of LNO.  A considerable degree of Fermi surface nesting leads to a susceptibility   $\chi$ ($q$) peaked close to the $q$ value at which the magnetic ordering has been observed in experiments.  Lee and co-workers noted that, for site-centered but not bond-centered spin-density waves, symmetry considerations imply that the magnetic ordering would lead also to a  charge ordering (with an amplitude of charge order proportional to  the square of the magnetic order parameter), providing a possible theoretical explanation of PrNiO$_3$ and NdNiO$_3$ for which the magnetic and charge ordering coincide. However, this  reasoning cannot be extended to  those nickelates  for which   the magnetic ordering and MIT were separated unless  unreasonably large values of $J_H$ are assumed. 

If  $\Delta$ is positive and reasonably large (but less than $U$) and $\beta$ (Eq.~\ref{Niconfig}) is still reasonably close to 1, one finds that the Mott-Hubbard picture provides a reasonable first-order description of the physics, with $\Delta$ playing the role of $U$.  This picture yields, for example, the basic understanding of the high-T$_c$ cuprates~ \cite{Zhang88,Medici09,Wang11}. However in the $RE$ nickelates,  $\Delta$ may be  negative, implying different physics. 
As noted above, DFT calculations indicate that the dominant electronic configuration is $d^8{\underline L}$ ($\gamma > \beta$) \cite{Han:2011p206804} and that  $\Delta$ is negative.  Mizokawa et al.,~\cite{Mizokawa00} explored the potential relevance of the negative charge transfer energy situation to nickelate physics  in the context of a Hartree-Fock analysis of an extended Hubbard model that also included oxygen orbitals. In this situation, holes are present on the oxygen, and a purely electronic mechanism for insulating behavior requires ordering of the oxygen holes. Ordering of oxygen holes  occurred only for extremely unphysically large magnitudes of  charge transfer energy ($\Delta \lesssim -7eV$), so these authors concluded that the combination of  breathing-mode distortions and strong correlations on the Ni site was essential for stabilizing the insulating state. 

Marianetti, one of us, and co-workers \cite{Han:2011p206804,PRL.109.156402,PRB.89.245133,PRB.90.235103} more quantitatively examined the negative charge-transfer energy  scenario  using a DFT+DMFT methodology  in which the correlated states were taken to be atomic-like Ni states constructed from Wannier functions defined over the full energy range of the Ni-O band complex and the O states were explicitly retained in the DMFT self-consistency equation.  As with all beyond-DFT approaches, the results depend to some degree on the values chosen for the interaction parameters $U$ and $J$ and for the double-counting correction \cite{Liechtenstein:95,Karolak10,Medici09,Wang11}. The important  results summarized here do not depend on the precise values chosen.  

The DFT+DMFT calculations lead to a mean \textit{d}-occupancy slightly smaller than  the $\sim 8.2$ value found in DFT, but still  closer to $8$ than to $7$. One recent calculation found  a mean value of $\sim 7.8$, \cite{PRL.109.156402,PRB.89.245133} with $\alpha^2=0.24$, $\beta^2=0.22$, $\gamma^2=0.43$ and  $\delta^2=0.11$ \cite{Parkprvt}. The dominance of the $d^8$ state is characteristic of all DFT+DMFT calculations based on atomic-like Ni orbitals. Furthermore, the physically relevant values of the Hund's couplings, $J_H \sim 0.7-1eV$ means that the only $d^8$ state that occurs with appreciable probability is the high-spin $d^8$ in which each $e_g$ orbital is occupied by one electron. This state is not susceptible to orbital polarization effects arising from strain or heterostructuring. In consequence, the orbital polarization is suppressed.  The  maximum polarization achievable under the strongest combinations of strain and heterostructuring is found to be at most  $\sim 25\%$~\cite{Chakhalian:2011p116805,Tung:2013p205112,Freeland:2011p57004,Benckiser:2011p189,Wu:2013p125124,Wu:2015p195130,PRB.90.045128}, far too small to provide the  one-band  cuprate-like physics anticipated on the basis of the Mott-Hubbard-like calculations.

By solving the DFT+DMFT equations in the paramagnetic phase using the atomic positions appropriate for insulating LuNiO$_3$ in its ground state, Park and co-workers \cite{PRL.109.156402} found a novel insulating state, which may be termed either a site-selective Mott state or a  hybridization wave  insulator. In this state, the mean \textit{d}-valences of the two structurally inequivalent Ni sites were almost the same, corresponding to approximately two   electrons in the $e_g$ orbitals. On each site, the two electrons are correlated by the Hunds coupling into an $S=1$ state. On the long-bond site, the $S=1$ is decoupled from the environment,  so the prediction is that, for example, an NMR measurement in the paramagnetic state would reveal a freely fluctuating $S=1$ local moment. On the other site, the $S=1$ is strongly hybridized to the two holes on the O-sites, leading to a singlet. The DFT+DMFT calculations provide an explicit realization of this state, which is insulating. As with other  DFT+ methodologies, the DFT+DMFT methodology indicates a ferrimagnetic ground state, rather than the long-period antiferromagnetic state observed experimentally. In the ferrimagnetic state, the moment size alternates: it is  small on the short-bond site and large on the long-bond site. These conclusions were subsequently reproduced by Johnson et al \cite{johnston} who performed exact diagonalization studies of the one- and three-dimensional Ni-O clusters.

More recently, Park et al.~\cite{PRB.89.245133,PRB.90.235103} performed total energy calculations within the DFT+DMFT methodology . By examining the variation of energy along a structural path interpolating from the cubic perovskite to the experimental LuNiO$_3$ structure these authors showed that the DFT+DMFT methodology reproduces the salient features of the pressure-$RE$ phase diagram of the materials, including the metallic nature of LNO and the critical pressures for the other compounds. DFT+$U$ methods were shown to greatly overestimate the critical pressures. 

These calculations establish  a reasonably consistent physical picture of the nickelates as materials with a modestly negative charge transfer energy, far from the Mott-Hubbard limit but with strong correlations, only a modest susceptibility to orbital polarization and an important coupling between electronic behavior and the lattice. The physics involves entangled Ni-$d$ and O$_{2p}$ states (with the entanglement crucially affected by the Hunds coupling).  An interesting challenge for future research is to understand the consequences of this negative-charge-transfer physics for other aspects of the behavior of the nickelates, including the magnetic order, the quasiparticle properties of metallic states, and dynamical electron-lattice effects.  In this regard, a very interesting step was   taken by Subedi et al. \cite{Subedi:2015p075128} who argued that the low energy physics of the negative-charge-transfer, Ni-O-entangled state may be represented by a negative $U$-positive $J_H$ Hubbard model. Further exploration of this and other theoretical approaches seems highly desirable.

\section{CONCLUDING REMARKS AND FUTURE ISSUES}

 With a brief introduction to the classification of   electronic structure for correlated oxides,  we   review above the recent progress in the field of interface-controlled ground-state engineering of the $RE$NiO$_3$ series and the simultaneous development of theoretical understanding about the origin of   transitions.  The possibility of realizing high-temperature superconductivity through orbital engineering was the initial motivation for studying \textit{RE}NiO$_3$  in ultra-thin heterostructures form, and the consequent efforts to stabilize various artificial geometries with unit cell precession have allowed  for the modulation of the ground states of the bulk phase  diagram.  Most importantly, this heterostructuring route has   also  resulted  in several unconventional phases such as insulating states without charge order, NFL metallic phases and antiferromagnetism in the metallic state, which are unattainable in the bulk.  Here, we  highlight  several important issues for future consideration. 

\subsection{Exploration of crystals beyond simple perovskites}
Although pushing synthesis of thin films of the highly distorted member of the series $RE$ = Y, Dy, Ho, ...Lu is an interesting avenue of research,  other geometries of Ni coordination  offer interesting possibilities for tuning the physical group states. As seen above, strained \textit{RE}NiO$_3$ films result in rather small orbital polarization and are limited  by the ability to distort the NiO$_6$ octahedra in the highly connected perovskite network. One effective way to induce larger orbital polarization could be  to move beyond simple perovskites into other classes of nickelates. One avenue involves layered nickelate phases in  Ruddlesden-Popper series\cite{Zhang:1994ux,Yoshizawa:2000dz} and  oxygen reduced cuprate-like variants $e.g.$  La$_3$Ni$_2$O$_6$~\cite{Poltavets:2009p046405}, La$_4$Ni$_3$O$_8$~\cite{Poltavets:2010p206403,Cheng:2012p236403}. The large tetragonal distortion or absence of the apical oxygen in these compounds gives rise to strong variations in the crystal field that can be used to tune orbital energies. In this series, thin film growth also offers opportunities to also tune orbitals via cation ordering~\cite{Nelson:2014p6884}. Another option, $A$NiO$_2$ compounds such as LiNiO$_2$ that display interesting electronic and magnetic states~\cite{Reynaud:2001ji,Kang:2007p195122}, remains largely unexplored from the perspective of epitaxial control.

\subsection{Role of oxygen in electronic and magnetic transitions}
Resonant X-ray absorption spectroscopy has firmly established the presence of holes on oxygen and the theoretical studies have also emphasized  the  ordering of the ligand holes  in a particular fashion across the transitions. Although the checker board type charge ordering  and the $E'$ magnetic structure of the Ni sublattice  have been firmly established by resonant X-ray scattering in bulk, the  ligand hole ordering and magnetism associated with the oxygen sub-lattice have not been experimentally verified yet due to  limitations of scattering geometry. An experimental answer to these crucial  issues can be the key to disentangling the puzzle of charge and   $E'$ magnetic ordering.  

\subsection{Ground State Engineering Using External Stimuli}
 
So far, this review  focuses around metastable states created using different architectures during growth. In this section, we explore the opportunities to drive these systems non-adiabatically even further away from equilibrium via external driving forces, and we focus on two areas: optical control with visible and mid-IR sources and ionic control by applying strong electrochemical gradients. Both approaches have unique portions of phase space that can be accessed as described below.

 \subsubsection{Ultrafast Dynamics in Nickelate Heterostructures}
Stimulating phase changes using optical pulses is a well-established way to control ground states in complex oxides~\cite{Zhang:2014dq} and offers the ability to drive a system far off the equilibrium axis as a means to trigger phase transitions or seek out wholly new emergent states. Initial work on nickelates focused on tracking the time response of the carrier dynamics~\cite{Katsufuji:1995:p4830,Ruello:2007jka} and coherent phonon excitation to probe the nature of electron-phonon coupling~\cite{Ruello:2009ew}, but recently this approach was extended using ultrafast soft X-rays to track the response of magnetism to optical excitation~\cite{Caviglia:2013to}. In NdNiO$_3$, the collapse of the antiferromagnetic order as  probed by resonant scattering at the AFM Bragg peak  revealed an ultrafast collapse of the ordering of the Ni moments, followed by  much slower dynamics for the parasitic magnetism related to low-temperature ordering of the Nd moments. Direct exploration of the magnetic order gives important insight into the interaction between the MIT and magnetism. However, the recovery process is much more difficult to understand due to the high energy of optical excitation, the system equilibrates on the longer time scale of several picoseconds due to heat generated during the carrier relaxation process via electron-lattice coupling. 

To circumvent this issue, researchers have been exploring mode-selective excitations that can efficiently transfer the energy in the optical excitation in to the system. One very intriguing approach  demonstrated recently  is the direct excitation of collective modes by light in the Thz-to-mid-IR regime~\cite{Forst:2011jx,Kampfrath:2013fl,Zhang:2014dq}. This approach was  first demonstrated in bulk complex oxides as a route to control the MIT by pumping specific phonon modes in the mid-IR that couple to the electronic degrees of freedom~\cite{Rini:2007hc} and  was later shown to be a  route to dynamically control the superconducting phase in cuprates~\cite{Fausti:2011dy}. Caviglia et al.~\cite{Caviglia:2012ea}   also applied this approach to nickelate films to control the MIT. The experiment  demonstrated  that,  by pumping  a substrate phonon mode that generates an ultra fast strain wave,   the system could be converted  to  the  metallic phase~\cite{Caviglia:2012ea}. By merging the mode-selective pumping with ultra fast soft X-ray scattering, Caviglia and colleagues~\cite{Forst:2015fv} were also able to measure the propagation of the phonon from the substrate into the film by following the spatial collapse of the magnetic state.  This  result confirmed  that the non-linear nature of the phonon excitation is a key component for triggering the transition~\cite{Forst:2011jx}. Very recently, a   DMFT theoretical framework has appeared to predict the couplings to design materials for far more efficient dynamic control~\cite{Subedi:2014ik}. In the future, this approach has the potential to offer new ways to efficiently manipulate correlated electron systems.

 \subsubsection{Ionic  Control}
Another new control avenue for nickelates is through the use of electric fields applied to thin-film samples to manipulate   carrier doping, analogous to   gate control in semiconducting devices~\cite{Ahn:2006dn}. Ionic liquids have emerged in this area as a means to apply large electric potentials through the use of a molecular electric double layer that forms on the surface of a sample. Although this approach was efficiently used to control the MIT in several  systems ~\cite{Ueno:2010vc,Asanuma:2010ti,Scherwitzl:2010fs,Nakano:2012jh,Shi:2014dh,Ha:2013p183102,Bubel:2015p122102}, the recent picture of charge carrier doping turned out to be more related to doping via the injection of oxygen vacancies under  large applied electrochemical potential gradients~\cite{Yang:2012is,Jeong:2013im}. Although  the timescale of this method is much slower due to doping via ionic diffusion, this process still offers the ability to change a material's phase through large stoichiometric changes, that were not available via gating in more conventional  oxide devices~\cite{Ahn:2006dn}. Recently, in another interesting approach,   the ionically gated \textit{RE }nickelates were used  to construct new classes of electronics based  on neuromorphic (i.e. brain-like) principles~\cite{Shi:2013gw,Ha:2014ek}.

\section*{ACKNOWLEDGMENTS}
 The work is supported primarily by the Indo-U.S. Joint Center for Rational Control of Functional Oxides under the Indo-U.S. Science and Technology Forum. S.M. was supported by the Department of Defense-Army Research Office grant 0402-17291, and J.C. was supported  by the Gordon and Betty Moore Foundation EPiQS Initiative through grant GBMF4534. Work at the Advanced Photon Source is supported by the U.S. Department of Energy, Office of Science under grant  DEAC02-06CH11357.  We thank S. Stemmer, S. Ramanathan, A. Georges, G. A. Fiete, and D. Khomskii for insightful comments.
 







%
%
%


\end{document}